\documentclass[twocolumn,pra,showkeys,a4paper]{revtex4-1}
\usepackage[latin1]{inputenc}
\usepackage{amsmath}
\usepackage{mathtools}
\usepackage{amsfonts}
\usepackage{amssymb}
\usepackage{graphicx}
\usepackage[caption=false]{subfig}
\usepackage{tikz}
\usepackage{parskip}
\usepackage{marginnote}
\usetikzlibrary{decorations.pathmorphing}
\usepackage{nomencl}
\makenomenclature

\usepackage{natbib}
\usepackage{booktabs}
\usepackage[export]{adjustbox}
\usepackage{soul}
 
\begin{document}
	\title{Effect of polydispersity on the phase behavior of non-additive hard spheres in solution, part II}
	
	\author{Luka Sturtewagen}
	\author{Erik van der Linden}%
	\affiliation{%
		Laboratory of Physics and Physical Chemistry of Foods,\\
		Wageningen University, Bornse Weilanden 9, 6708 WG Wageningen, The Netherlands 
	}%

	\begin{abstract}
		We study the theoretical phase behavior of an asymmetric binary mixture of hard spheres, of which the smaller component is monodisperse and the larger component is polydisperse. The interactions are modeled in terms of the second virial coefficient and can be additive hard sphere (HS) or non-additive hard sphere (NAHS) interactions. The polydisperse component is subdivided into two sub-components and has an average size ten or three times the size of the monodisperse component. We give the set of equations that defines the phase diagram for mixtures with more than two components in a solvent. We calculate the theoretical liquid-liquid phase separation boundary for two phase separation (the binodal) and three phase separation, the plait point, and the spinodal. We vary the distribution of the polydisperse component in skewness and polydispersity, next to that we vary the non-additivity between the sub-components as well as between the main components. We compare the phase behavior of the polydisperse mixtures with binary monodisperse mixtures for the same average size and binary monodisperse mixtures for the same effective interaction. We find that when the compatibility between the polydisperse sub-components decreases, three-phase separation becomes possible. The shape and position of the phase boundary is dependent on the non-additivity between the subcomponents as well as their size distribution. We conclude that it is the phase enriched in the polydisperse component that demixes into an additional phase when the incompatibility between the sub-components increases.
	\end{abstract}
	
	\maketitle
	\section{Introduction}
	In the study of the phase behavior of binary mixtures, the components are usually assumed to be pure and monodisperse, however in nature most components are not that neatly monodisperse. Many components show size and charge variation or contain hard to remove particles that can influence their phase behavior in binary mixtures. In their experimental work, Sager \cite{Sager1998} reported that even small impurities can lead to drastic shifts in the position of the phase boundary. Next to that the compatibility between components can be depended on the temperature \cite{Edelman2001}, salt concentration or pH of the solution\cite{Kontogiorgos2009}. 
	
	Two different physical mechanisms drive the phase separation between hard spheres. The first one involves only excluded volume interactions. In this mechanism the minimal distance between the particles is determined by the sum of their respective radii \cite{Biben1997}. This is the typical additive hard sphere interaction (HS). With this mechanism, phase separation is driven by a size asymmetry between the particle sizes \cite{Biben1991}. This asymmetry leads to depletion of small spheres around the large spheres and as a result to an effective attraction (depletion interaction) between the larger spheres \cite{Dijkstra1999}. The other mechanism is when the distance between the particles of a different species can be larger or smaller than the sum of their respective radii. This is referred to as non-additive hard sphere (NAHS) interaction. Previous research has shown that already at small degrees of non-additivity it becomes possible for components with no size asymmetry to demix \cite{Roth2001} \cite{Dijkstra1998}. Either way, upon phase separation, the mixture will demix into two (or more) phases, each enriched in one of the components. In the previous article, we focused on the first type of interaction \cite{Sturtewagen2019}. We investigated the influence of size polydispersity on the phase behavior of an additive binary asymmetric mixture. In this work we will focus on the second type, binary (polydisperse) mixtures where the distance between the particles of different species can be larger or smaller than the sum of their respective radii. 
	
	Piech and Walz \cite{Piech2000} studied the effect of size polydispersity and charge heterogeneity on the depletion interaction in a colloidal system. They found that the size distribution in the larger particle had a different effect on the depletion attraction for charged and non-charged hard sphere systems. For the depletion attraction decreased between the larger particles at constant volume fraction due to the polydispersity. This effect was further enhanced by the presence of charge. Polydispersity significantly lowers the magnitude of the repulsive barrier.
	
	The non-additivity is usually described by the non-additivity parameter $\Delta$ (with $\Delta \geq -1$). When $\Delta = 0$ the mixture has additive hard sphere interaction and the closest approach of the particles is the sum of their radii. When $\Delta < 0$ the two particles experience more attraction and can come closer to each other than the sum of their radii, while when $\Delta > 0$ the two particles have more repulsion and their distance of closest approach is larger than the sum of their respective radii. It is clear that this can have enormous effects on their phase behavior. Particles with a negative $\Delta$ tend to be more compatible with each other, while particles with a positive $\Delta$ are less compatible and tend to demix at lower concentrations. Already at the relatively low $\Delta = 0.1$, it becomes possible for components with the same size to demix \cite{Sillren2010}.
	
	Paricaud \cite{Paricaud2008} studied the phase behavior of polydisperse colloidal dispersions. Their mixture consisted of a monodisperse component and a polydisperse component. The interaction between the monodisperse and polydisperse components was assumed to be NAHS (with the same $\Delta$ for all polydisperse sphere), while the interaction between the polydisperse components amongst themselves was assumed to be additive HS. They find that the critical point of a polydisperse mixture is at lower solution pressure than for completely monodisperse mixtures. For mixtures with large variation in the size of the polydisperse mixtures they observe the possibility of a three phase system. The phase behavior of a colloid and a polydisperse polymer was studied by \cite{Sear1997}. They used the Asakura-Osawa model fo the interactions between the different components. They found that increasing the polydispersity increased also the extent of the fluid-fluid coexistence. They reason that the introduction of larger polymer coils is the driving force towards phase separation.
		
	In this study we aim to get a better understanding of how non-additive interaction influences the phase behavior of binary mixtures with some polydispersity or impurities. We will study the position of the phase separation boundary, the spinodal, and the critical point. Next to that we aim to predict the fractionation of the polydisperse component between the different phases. We model the interactions between the different components using the second virial coefficient (\ref{ss:Virial}). In section \ref{ss:stability} we describe the equations for the spinodal, in section \ref{ss:cp} we describe the equations for the critical point and finally in section \ref{ss:phase} we describe the equations defining the phase boundary. With the expressions in section \ref{s:Theory} we have enough to calculate the phase diagram for a variety of mixtures described in section \ref{s:RD}. First we introduce non-additivity between the main components in the binary mixtures (\ref{ss:DeltaAB}), subsequently we introduce non-additivity between the sub-components in the polydiseperse component (\ref{ss:DeltaBB}), and finally we combine both in section \ref{ss:DeltaBBDeltaAB}. In section \ref{ss:Fractionation} we look into the fractionation of some of the mixtures from \ref{ss:DeltaBB} at a specific parent concentration.
	
	\section{\label{s:Theory}Theory}
	We show the equations used for the calculations of the phase diagram of the different studied systems: the set of equations defining the stability boundary, the critical point, and phase boundaries of a mixture. All sets of equations are solved in Matlab R2017b. For a more detailed derivation of the equations, we refer to \cite{Sturtewagen2019}. 
	
	\subsection{\label{ss:Virial}Osmotic virial coefficient}
	The osmotic pressure, $\Pi$, of a solution at a temperature $T$, can be written as a virial expansion, similar to the virial expansion of the universal gas law for real gasses \cite{Hill1986}:
	\begin{equation}
	\beta\Pi = \rho + B_2(T,\mu_s)\rho^2 + B_3(T,\mu_s)\rho^3 + ... \label{eq:EquationofState}
	\end{equation}
	\nomenclature{$\Pi$}{Osmotic pressure}
	\nomenclature{$T$}{Temperature}
	\nomenclature{$\rho$}{Density $\displaystyle\frac{N}{V}$}
	\nomenclature{$B$}{Virial coefficient}
	\nomenclature{$B_2$}{Second virial coefficient}
	\nomenclature{$\mu_s$}{Chemical potential of the solution}
	\nomenclature{$N$}{Number of particles}
	\nomenclature{$V$}{Volume}
	\nomenclature{$N_{\nu}$}{Number of solute particle}
	\nomenclature{$k$}{Boltzmann's constant}
	\nomenclature{$\beta$}{$\displaystyle\frac{1}{kT}$}
	\nomenclature{$\nu_i$}{Compound of type $\nu_i$}
	\nomenclature{$n$}{Number of distinguishable components}
	
	with $\displaystyle\beta = \frac{1}{kT}$, $k$ the Boltzmann's constant, $\rho$ the number density of the component $\displaystyle\left(\frac{N_\nu}{V}\right)$, $B_2$ the second virial coefficient, and $B_3$ the third virial coefficient. The second virial coefficient accounts for the increase in osmotic pressure due to particle pairwise interaction. The third virial coefficient accounts for the interaction between three particles in a variety of configurations. The equation can be expanded for higher densities with $B_n$, the $n^{th}$ virial coefficient, which accounts for the interaction between $n$ different particles.
	
	In this work we will limit the virial expansion to the second virial coefficient, which is given by \cite{Lekkerkerker2011}:	
	\begin{equation}
	B(T,\mu_s) = 2\pi\int_{0}^{\infty}r^2(1-\exp{[-\beta W(r)]})dr \label{eq:2ndvirial}
	\end{equation}
	\nomenclature{$W(r)$}{Interaction potential}
	in which $\mu_s$ is the chemical potential of the solution, $W(r)$ is the interaction potential between the particles, and $r$ is the distance.
	
	For additive hard sphere (HS) interaction, the interaction potential for two particles (of the same species or different species) is given by:
	\begin{equation}
	W(r)_{HS} = \left\{
	\begin{array}{ll} 0, & r > \sigma_{ij} \\
	\infty, & r \leq \sigma_{ij}
	\end{array}\right. \label{eq:HScrosspotential}
	\end{equation}
	\nomenclature{$\sigma$}{Diameter of a particle}
	with $\sigma_{ij} = (\sigma_i + \sigma_j)/2$, the distance between the centers of the two particles. 
	
	For non-additive hard spheres (NAHS), the distance of the closest approach of the centers of the two particles of different species can be closer or further than the distance between their centers \cite{Roth2001}. The closest distance then becomes: $\sigma_{ij} = ((\sigma_i + \sigma_j)/2)(1+\Delta)$, in which $\Delta$ $(\geq -1)$ accounts for the non-additivity of the interaction between the particles. When $\Delta > 0$ the distance of closest approach of both spheres increases and when $\Delta < 0$ the distance of closest approach decreases compared to that due to HS interaction only. For additive hard sphere interaction $\Delta = 0$.
	
	In a mixture with $n$ distinguishable components in a solution, there are two main types of two particle interactions that can occur: between particles of the same species and particles of different species.
	
	For the second virial coefficient given by eq. \ref{eq:2ndvirial}, using the interaction potential defined in eq. \ref{eq:HScrosspotential}, we find:
	\begin{align}
	B_{xx} &= \frac{2\pi}{3}(\sigma_x)^3 \label{eq:B_xx}\\
	B_{xy} &= \frac{2\pi}{3}\left(\left(\frac{\sigma_x+\sigma_y}{2}\right)(1+\Delta)\right)^3 \label{eq:B_xy}
	\end{align}
	\nomenclature{$\Delta$}{Non-additivity parameter for interacting particles}
	\nomenclature{$x$}{Particle of type $x$}
	\nomenclature{$y$}{Particle of type $y$}
	where $B_{xx}$ is the second virial coefficient for particles of the same species (assumed to be HS) and $B_{xy}$ is the second virial coefficient of particles of different species, which can be HS or NAHS.
	
	The general equation for the osmotic pressure for a dilute mixture is given by \cite{Sturtewagen2019}:
	\begin{align}
	\beta \Pi &= \rho + B_{11}\rho_{1}^2 + 2B_{12}\rho_{1} \rho_{2} + 2B_{13}\rho_{1} \rho_{3}  ... \nonumber\\
	&= \rho + \sum_{i}^{n}\sum_{j}^{n}B_{ij}\rho_{i}\rho_{j} +... \label{eq:OsmoticGeneral}
	\end{align}
	
	In this article we focus on binary mixtures in which one of the components consists of sub-components. By increasing the number of sub-components, the number of equations to solve for the phase diagram increases. Just as in article \cite{Sturtewagen2019} we also compare the results to the number averaged virial coefficients of the different components. The number averaged virial coefficient was chosen because it allows for comparison to experiments, e.g. the virial coefficient obtained from osmometric measurements \cite{Ersch2016}.
	
	The number averaged second virial coefficient of a mixture can be written as:
	\begin{equation}
	\begin{aligned}
	B_{mix} &= B_{11}x_{1}^2 + 2B_{12}x_{1} x_{2} + 2B_{13}x_{1} x_{3}  ... \\
	&= \sum_{i}^{m}\sum_{j}^{m}B_{ij}x_{i}x_{j} \label{eq:Bmix}
	\end{aligned}
	\end{equation}
	in which $B_{ii}$ is the second virial coefficient of the $i$\textsuperscript{th} particle, $B_{ij}$ is the second cross virial coefficient of the $i$\textsuperscript{th} particle and the $j$\textsuperscript{th} particle, and $x_{i}$ is the fraction of the $i$\textsuperscript{th} particle, $\displaystyle\sum x_{i}=1$.
	
	Using this definition, we can map the binary mixture consisting of for example a monodisperse component 1 and a component 2 subdivided into two subcomponents ($a$ and $b$) by a $2\times2$ matrix of virial coefficients. We will refer to this $2\times 2$ matrix as the effective virial coefficient matrix.
	\begin{eqnarray}
	B_{11_{eff}} & = & B_{11} \nonumber\\
	B_{12_{eff}} & = & x_aB_{12_a} + x_bB_{12_b}\nonumber\\
	B_{22_{eff}} & = & x_a^2B_{2_a2_a} + 2x_ax_bB_{2_a2_b}+x_b^2B_{2_b2_b}\label{eq:Beff2}
	\end{eqnarray}
	The effective virial coefficient matrix for this mixture becomes then:
	\begin{equation}
	B_{eff}=\left[\begin{matrix}
	B_{11_{eff}} &  B_{12_{eff}}\\[3ex]
	B_{12_{eff}} &  B_{22_{eff}}
	\end{matrix}\right]
	\end{equation}
	
	\subsection{\label{ss:stability}Stability of a mixture}
	The stability of a mixture is dependent on the second derivative of the free energy. If the second derivative of the mixture becomes zero, the mixture is at the boundary of becoming unstable. Unstable mixtures have a negative second derivative \cite{Heidemann1975} \cite{Beegle1974}.
	
	The differential of the free energy of a mixture is given by \cite{Hill1986}:
	
	\begin{equation}
	d A = -S d T - p dV + \sum_i^n \mu_{i} d N_{i} \label{eq:Helmholtzdifferential}
	\end{equation}
	\nomenclature{$A$}{Helmholtz free energy}
	\nomenclature{$p$}{Pressure}
	\nomenclature{$\mu_i$}{Chemical potential of component $i$}
	\nomenclature{$S$}{Entropy}
	in which $\mu_{i}$, the chemical potential (the first partial derivative of the free energy with respect to number of particles ($N_i$)) for component $i$ is given by:
	\begin{equation}
	\mu_i = \mu_i^0 + kT \ln(\rho_i)+2kT\left(\sum_{j}^{n} B_{ij}\rho_j\right) \label{eq:Chempotential}
	\end{equation}
	For a mixture with $n$ distinguishable components, the second partial derivatives can be represented by a $n\times n$ matrix of the first partial derivatives of the chemical potential of each component.
	
	This results in the following general stability matrix:
	\begin{align}
	M_1 &= \begin{bmatrix}\displaystyle
	\frac{\partial\mu_1}{\partial N_1} & \cdots & \displaystyle\frac{\partial\mu_1}{\partial N_n} \\
	\vdots & \ddots & \vdots \\
	\displaystyle\frac{\partial\mu_n}{\partial N_1} & \cdots & \displaystyle\frac{\partial\mu_n}{\partial N_n}
	\end{bmatrix} \nonumber\\ 
	&= 
	\begin{bmatrix}
	2B_{11}+\displaystyle\frac{1}{\rho_1} & \cdots & 2B_{1n} \\
	\vdots & \ddots & \vdots \\
	2B_{1n} & \cdots & 2B_{nn} + \displaystyle\frac{1}{\rho_n}
	\end{bmatrix} \label{eq:Firstcriterion}
	\end{align}
	The mixture is stable when all eigenvalues are positive \citep{Solokhin2002}, when on the other hand one of the eigenvalues is not positive, the mixture becomes unstable. The limit of stability is reached when the matrix has one zero eigenvalue and is otherwise positive definite, and is referred to as the spinodal \citep{Heidemann1980}.
	
	When there are only two components in the mixture ($n=2$), the spinodal is defined by the condition $\det{M_1} =0$. When the number of components is larger ($n>2$), $\det{M_1} =0$ can have more than one solution \citep{Solokhin2002}. The spinodal can be found by checking whether $\det{M_1}$ changes sign for small changes in the concentrations of the components.
	
	\subsection{\label{ss:cp}Critical points}
	In a binary mixture, the critical point is a stable point which lies on the stability limit (spinodal) \cite{Heidemann1980} and where the phase boundary and spinodal coincide. In mixtures of more components these critical points become plait points. Critical points and plait points are in general concentrations at which two phases are in equilibrium and become indistinguishable \cite{Heidemann2013}.
	
	There are two criteria that can be used to find critical points. The first one is $det(M_1)=0$, which is the equation for the spinodal. The other criterion is based on the fact that at the critical point, the third derivative of the free energy should also be zero. For a multicomponent system, this criterion can be reformulated using Legendre transforms as $det(M_2) = 0$ \cite{Beegle1974}\cite{Reid1977}, where:
	\begin{equation}
	M_2=\begin{bmatrix}
	\displaystyle\frac{\partial\mu_1}{\partial N_1} & \cdots & \displaystyle\frac{\partial\mu_n}{\partial N_n} \\
	\vdots & \ddots & \vdots \\
	\displaystyle\frac{\partial M_1}{\partial N_1} & \cdots & \displaystyle\frac{\partial M_1}{\partial N_n}
	\end{bmatrix} \label{eq:Secondcrit}
	\end{equation}
	Matrix $M_2$ is matrix $M_1$ with one of the rows replaced by the partial derivatives of the determinant of matrix $M_1$. 
	Note: it does not matter which row of the matrix is replaced. 
	
	\subsection{\label{ss:phase}Phase boundary}
	When a mixture becomes unstable and demixes into two or more phases, the chemical potential of each component and the osmotic pressure is the same in all phases \cite{Hill1986}.
	\begin{equation}
	\left\{
	\begin{aligned}
	\beta\Pi^{I} &= \beta\Pi^{II}  &=\cdots\\
	\beta\mu_{1}^{I} &= \beta\mu_{1}^{II} &=\cdots\\
	&\vdots&\\
	\beta\mu_{n}^{I} &= \beta\mu_{n}^{II} &=\cdots\\ 
	\end{aligned}
	\right.
	\end{equation}
	where the phases are denoted by $I,II,...$.
	
	For a mixture containing $n$ distinguishable components, that demixes into two phases, this results in $n+1$ equations and $2\times n$ unknowns. If the mixture demixes into three phases, this results in $2 \times n + 2$ equations and $3\times n$ unknowns. To solve the set of equations without having to fix the concentration of one component and the ratio between the other components for at least one of the phases, we need extra equations. For the extra set of equations, we build on the fact that no particles are lost and no new particles are created during phase separation, and the fact that we assume that the total volume does not change.
	
	For a system that separates into three phases we then obtain:
	\begin{equation}
	\rho = \sum_i^n \rho_i = \frac{\sum\limits_i^n N_i}{V} = \frac{\sum\limits_i^n N_i^{I}+\sum\limits_i^n N_i^{II}+\sum\limits_i^n N_i^{III}}{V^{I}+V^{II}+V^{III}} \nonumber
	\end{equation}
	which can be rewritten as \cite{Sturtewagen2019}:
	\begin{align}
	\rho &=\alpha^{I} \sum_i^{n}\rho_i^{I} + \alpha^{II} \sum_i^{n}\rho_i^{II} + (1-\alpha^{I}-\alpha^{II}) \sum_i^{n}\rho_i^{III}\nonumber
	\end{align}
	with
	\begin{equation}
	\alpha^{I} = \frac{V^{I}}{\sum\limits_i^{f}V^{i}} \nonumber
	\end{equation}
	in which $f$ denotes the number of phases.
	
	This results in the following set of equations:
	\begin{equation}
	\left\{
	\begin{aligned}
	\beta\Pi^{I} &= \beta\Pi^{II}  =\cdots\\
	\beta\mu_{1}^{I} &= \beta\mu_{1}^{II} =\cdots\\
	&\vdots\\
	\beta\mu_{n}^{I} &= \beta\mu_{n}^{II} =\cdots\\ 
	\rho_1 &= \alpha^{I} \rho_1^{I} + \cdots + \left(1-\sum^{f-1}_i\alpha^{i}\right) \rho_1^{f} \\
	&\vdots\\
	\rho_n &= \alpha^{I} \rho_n^{I} + \cdots + \left(1-\sum^{f-1}_i\alpha^{i}\right) \rho_n^{f}
	\end{aligned}
	\right.
	\end{equation}
	With this set of equations, we have $2\times n +1$ unknowns and $2\times n +1$ equations for mixtures that separate into two phases. For mixtures that demix into three phases, we have $3\times n + 2$ unknowns and $3\times n+2$ equations. Therefore, this set of equations allows for calculating the concentration of each component in each of the phases for any given parent concentration, given that the mixture will demix at this concentration.
	
	\section{\label{s:RD}Results and discussion}
	
	In this work we calculated the phase diagram for a variety of binary non-additive mixtures of a small hard sphere $A$ and a larger hard sphere $B$ with a size ratio $q = \sigma_A / \sigma_B$. Component $B$ is sub-divided into two sub-components and is characterized by a degree in polydispersity ($PD$), defined by:
	\begin{equation}
	PD = \frac{\sqrt{\sum{(\sigma_{B_i} - \sigma_B)^2}\times N_{B_i}/N_B}}{\sigma_B}\times 100\nonumber 
	\end{equation}
	\nomenclature{$q$}{Size ratio between components $\displaystyle\frac{\sigma_A}{\sigma_B}$}
	\nomenclature{$\eta$}{Packing fraction $\displaystyle\frac{\pi \rho \sigma^3}{6}$}
	\nomenclature{$PD$}{Degree in polydispersity}
	We varied the non-additivity between particles of component $A$ and $B$ ($\Delta_{AB}$), and between the sub-components of $B$ ($\Delta_{B_aB_b}$). Next to that we varied the degree of polydispersity ($PD$) of component $B$ and the distribution between the sub-components as well as the size ratio ($q$) between component $A$ and $B$.
	
	For all particles, the concentrations are expressed as a dimensionless parameter according to $\displaystyle \eta = \frac{\pi \rho \sigma^3}{6}$. We calculated the critical point, the phase separation boundary, and the spinodal of the various mixtures. Next to that, we also investigated the composition of the child phases, volume fraction of the phases ($\alpha$), and the fractionation of the polydisperse component $B$ for a specific parent mixture ($\eta = (0.010,0.200)$), for mixtures with a size ratio $q = \sigma_A / \sigma_B = 1/10$ and $\Delta_{AB} = 0$, while varying the non-additive interaction between the sub-components of $B$ ($\Delta_{B_aB_b}$). 
	
	\subsection{\label{ss:DeltaAB} Non-additive interaction between component $A$ and $B$ ($\Delta_{AB}$)}
	
	\begin{figure}
		\centering
		\includegraphics[trim=1cm 0cm 2cm 0.5cm, clip=true ,width=0.47\textwidth]{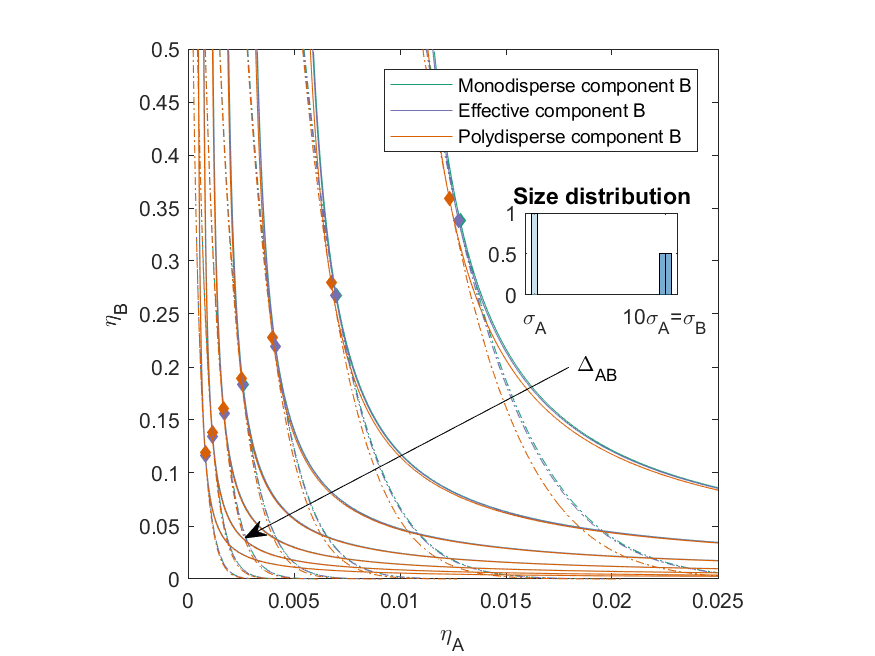}
		\caption{\label{Fig:narrow5050DeltaAB}Phase diagram for binary (component $A$ and $B$) non-additive hard sphere mixture with size ratio $q = \sigma_A / \sigma_B =  1/10$, component $A$ is monodisperse, component $B$ is polydisperse ($PD = 4.00$), plotted as a function of the partial packing fractions, $\eta_A$ and $\eta_B$. The interaction between components $A$ and $B$ is non-additive, the non-additivity parameter $\Delta_{AB}$ was varied from -0.1 to 0.5 with a step size of 0.1 (the arrow indicates increasing $\Delta_{AB}$). The interaction between the sub-components $B$ is additive. The spinodal (solid line) and binodal (dashed line) meet each other at the critical point (diamond).}
	\end{figure}

	\begin{figure}
		\includegraphics[trim=1cm 0cm 1cm .5cm, clip=true ,width=0.48\textwidth]{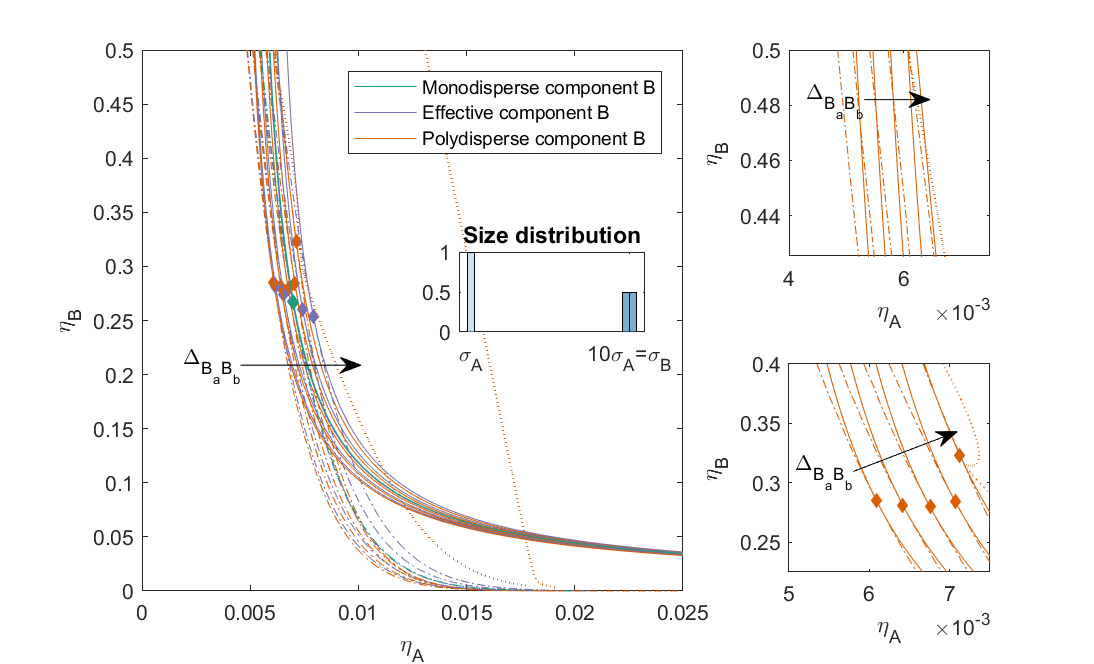}
		\caption{\label{Fig:narrow5050DeltaBB}Phase diagram for binary (component $A$ and $B$) non-additive hard sphere mixture with size ratio $q = \sigma_A / \sigma_B =  1/10$, component $A$ is monodisperse, component $B$ is polydisperse ($PD = 4.00$), plotted as a function of the partial packing fractions, $\eta_A$ and $\eta_B$. The interaction between components $A$ and $B$ is additive, the interaction between the sub-components $B$ is non-additive, the non-additivity parameter $\Delta_{B_aB_b}$ was varied from -0.1 to 0.1 with a step size of 0.05 (the arrow indicates increasing $\Delta_{B_aB_b}$). The spinodal (solid line) and binodal (dashed line) meet each other at the plait point (diamond), the three phase boundary is indicated with a dotted line and meets the spinodal at the plait point (diamond)}
	\end{figure}

	\begin{figure*}
		\centering
		\subfloat[\label{Fig:PD8}$PD = 8.00$]{%
			\includegraphics[trim=1cm 0cm 1cm .5cm, clip=true ,width=0.49\textwidth]{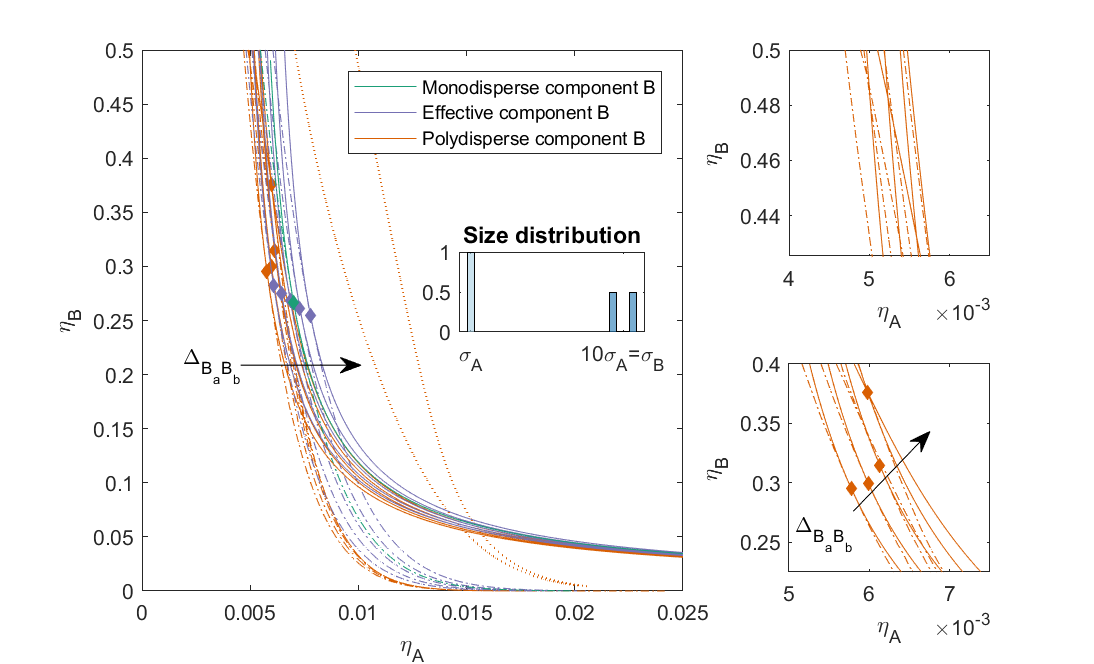}%
		}
		\centering 
		\subfloat[\label{Fig:PD12}$PD = 12.00$]{%
			\includegraphics[trim=1cm 0cm 1cm .5cm, clip=true ,width=0.49\textwidth]{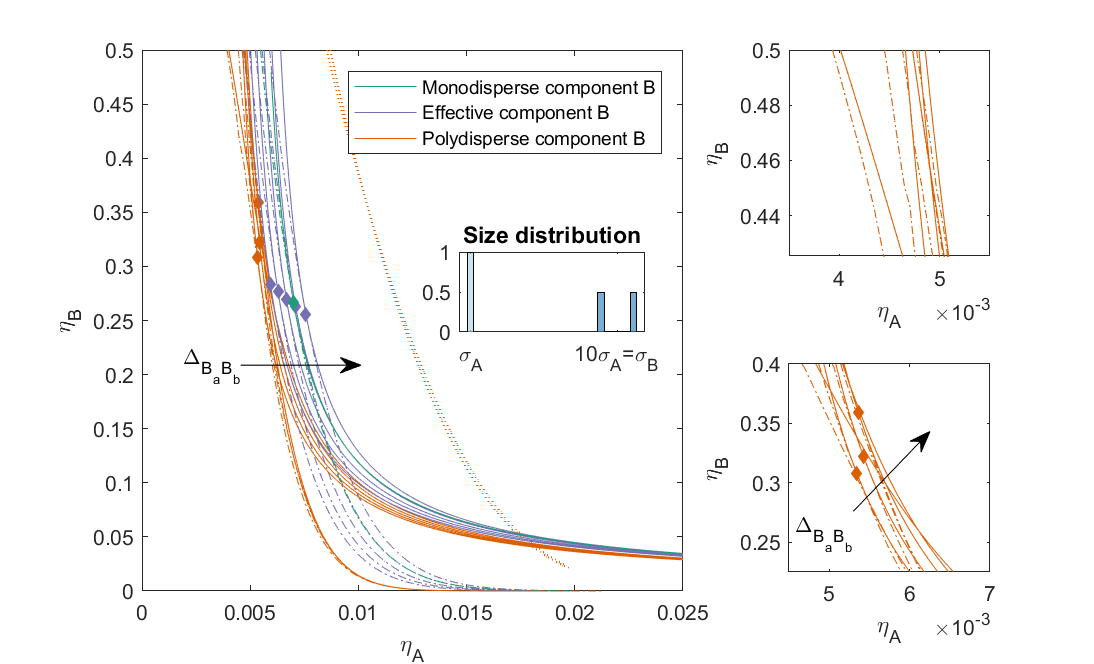}%
		}
		\caption{\label{Fig:broader5050DeltaBB}Phase diagram for binary (component $A$ and $B$) non-additive hard sphere mixture with size ratio $q = \sigma_A / \sigma_B =  1/10$, component $A$ is monodisperse, component $B$ is polydisperse ($PD = 8.00$ or $PD = 12.00$), plotted as a function of the partial packing fractions, $\eta_A$ and $\eta_B$. The interaction between components $A$ and $B$ is additive, the interaction between the sub-components $B$ is non-additive, the non-additivity parameter $\Delta_{B_aB_b}$ was varied from -0.1 to 0.1 with a step size of 0.05 (the arrow indicates increasing $\Delta_{B_aB_b}$). The spinodal (solid line) and binodal (dashed line) meet each other at the plait point (diamond), the three phase boundary is indicated with a dotted line.}
	\end{figure*}
	
	For the first set of mixtures (see figure \ref{Fig:narrow5050DeltaAB}), we calculated the phase diagram for binary mixtures with non-additive interaction between monodisperse component $A$ and slightly polydisperse component $B$, with two sub-components and a $PD = 4.00$. These two sub-components are additive hard spheres in two sizes (both present in the same amount), with the number average size of the mixture equal to 10 times the size of component $A$. The mixture therefore consist of three components of different size. We varied the non-additivity between component $A$ and $B$ ($\Delta_{AB}$, the same for both sub-components) from -0.1 to 0.5 with a step size of 0.1. When $\Delta_{AB} = 0$, the interaction between all components equals additive hard sphere interaction. We calculated the phase diagram using both the simplified $2 \times 2$ effective virial coefficient matrix described in the theory (we refer to this as the effective mixture $B$) and the full $3 \times 3$ virial coefficient matrix (to which we refer as the polydisperse mixture $B$). These mixtures are also compared to mixtures in which component $B$ is monodisperse with a size equal to the average particle size of component $B$ (we refer to this as the monodisperse mixture $B$). 
	
	With increasing $\Delta_{AB}$, the phase boundary, spinodal and critical point shifts towards lower concentrations, for the monodisperse mixture, effective mixture, and polydisperse mixture. This is in line with research on non-additive binary mixtures \cite{Hopkins2010}. The difference between the phase boundary, spinodal and critical point of the monodisperse mixture and the effective mixture is negligible, for all $\Delta_{AB}$. We see however that the introduction of the polydispersity causes the critical point to shift to a higher volume fraction of component $B$ and that especially at lower volume fraction of component $B$ the phase separation boundary shifts towards slightly lower packing fractions. This effect is more pronounced when $\Delta_{AB}$ is small. 
	
	When the $PD$ of component $B$ is increased, or the distribution of the sub-components of $B$ is varied, we see the same pattern as in figure \ref{Fig:narrow5050DeltaAB} (see supplementary materials). However, we see that, just like discussed in \cite{Sturtewagen2019}, the critical point shifts towards higher concentrations of $B$ for the polydisperse mixtures depending on the size and concentration of the largest sub-component of $B$ and the difference between the effective and the monodisperse mixtures increases with the size of the largest sub-component of $B$. 

	\begin{figure*}
		\centering
		\subfloat[\label{Fig:Left25}Left skewed]{%
			\includegraphics[trim=1cm 0cm 1cm .5cm, clip=true ,width=0.49\textwidth]{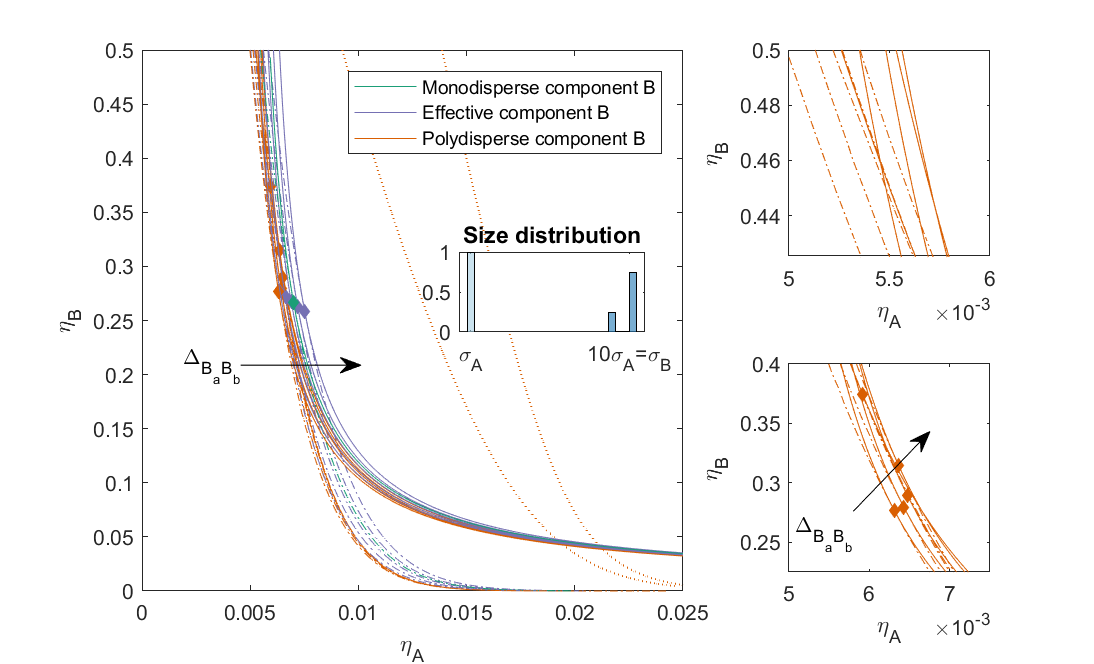}%
		}
		\centering 
		\subfloat[\label{Fig:Right25}Right skewed]{%
			\includegraphics[trim=1cm 0cm 1cm .5cm, clip=true ,width=0.49\textwidth]{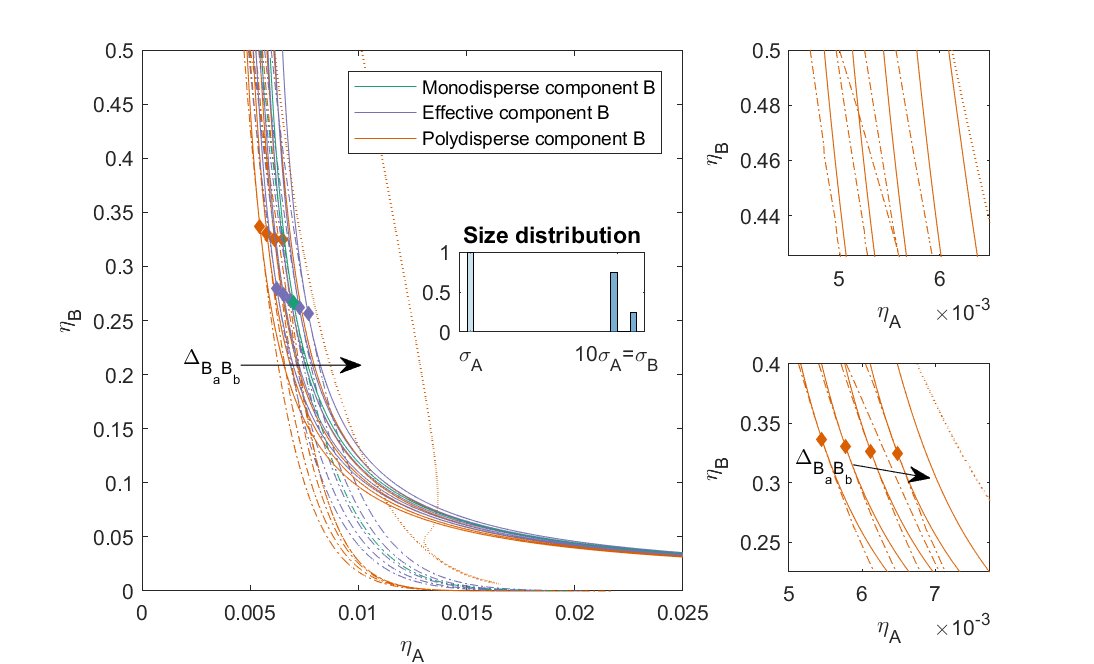}%
		}
		\caption{\label{Fig:PD2575}Phase diagram for binary (component $A$ and $B$) non-additive hard sphere mixture with size ratio $q = \sigma_A / \sigma_B =  1/10$, component $A$ is monodisperse, component $B$ is polydisperse ($PD = 6.93$), plotted as a function of the partial packing fractions, $\eta_A$ and $\eta_B$. The interaction between components $A$ and $B$ is additive, the interaction between the sub-components $B$ is non-additive, the non-additivity parameter $\Delta_{B_aB_b}$ was varied from -0.1 to 0.1 with a step size of 0.05 (the arrow indicates increasing $\Delta_{B_aB_b}$). The spinodal (solid line) and binodal (dashed line) meet each other at the plait point (diamond), the three phase boundary is indicated with a dotted line.}
	\end{figure*}

	\begin{figure*}
		\centering
		\subfloat[\label{Fig:Left10}Left skewed]{%
			\includegraphics[trim=1cm 0cm 1cm .5cm, clip=true ,width=0.49\textwidth]{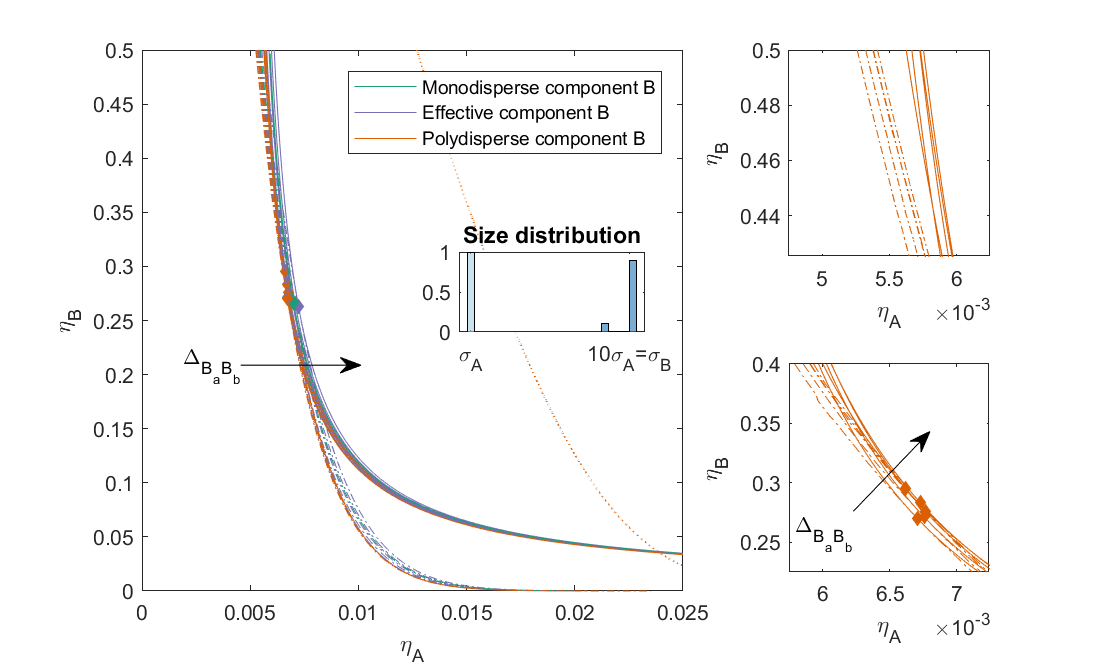}%
		}
		\centering 
		\subfloat[\label{Fig:Right10}Right skewed]{%
			\includegraphics[trim=1cm 0cm 1cm .5cm, clip=true ,width=0.49\textwidth]{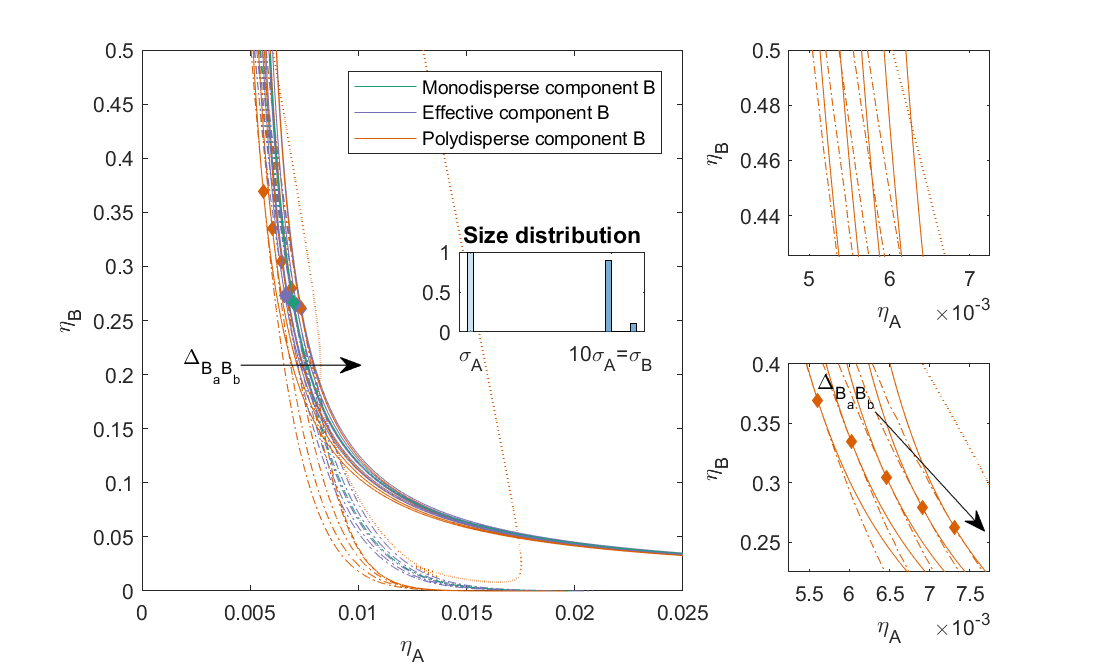}%
		}
		\caption{\label{Fig:PD1090}Phase diagram for binary (component $A$ and $B$) non-additive hard sphere mixture with size ratio $q = \sigma_A / \sigma_B =  1/10$, component $A$ is monodisperse, component $B$ is polydisperse ($PD = 4.80$), plotted as a function of the partial packing fractions, $\eta_A$ and $\eta_B$. The interaction between components $A$ and $B$ is additive, the interaction between the sub-components $B$ is non-additive, the non-additivity parameter $\Delta_{B_aB_b}$ was varied from -0.1 to 0.1 with a step size of 0.05 (the arrow indicates increasing $\Delta_{B_aB_b}$). The spinodal (solid line) and binodal (dashed line) meet each other at the plait point (diamond), the three phase boundary is indicated with a dotted line.}
	\end{figure*}

	\subsection{\label{ss:DeltaBB} Non-additive interaction within polydisperse component $B$ ($\Delta_{B_aB_b}$)}
	In the next set of mixtures, we kept the interaction between the components $A$ and $B$ as hard-sphere additive, but we introduced some non-additivity in the interaction between the sub-components of $B$. we varied $\Delta_{B_aB_b}$ from $-0.10$ to $0.10$ with a step size of $0.05$. When $\Delta_{B_aB_b}$ is small, the sub-components are more compatible with each other, when on the other hand $\Delta_{B_aB_b}$ increases and becomes positive, the compatibility between the sub-components decreases. When $\Delta_{B_aB_b} > 0$ it becomes possible for components of similar size to phase separate \cite{Hopkins2010}. 

	In figure \ref{Fig:narrow5050DeltaBB} we plotted the phase diagram for the binary mixtures with $PD = 4.00$, $\Delta_{AB} = 0$ and we varied $\Delta_{B_aB_b}$. When $\Delta_{B_aB_b} > 0$, the compatibility between the sub-components decreases and phase separation into three phases becomes possible (depicted as the dotted line in the figure). Mixtures with a smaller $\Delta_{B_aB_b}$ demix into two phases  at lower concentrations compared to the completely hard sphere mixture. Mixtures with a larger $\Delta_{B_aB_b}$ demix into two phases at higher packing fractions compared to the completely hard sphere mixture, and also have a three phase boundary. The three-phase boundary is at lower concentrations for larger $\Delta_{B_aB_b}$ and comes close to the two-phase boundary for the mixture with $\Delta_{B_aB_b} = 0.10$. 
	The critical point of the polydisperse mixtures changes depending on the non-additivity of the sub-components: The critical point is at its lowest concentrations of $A$ for negative $\Delta_{B_aB_b}$, its lowest concentration of $B$ is when the interaction between the sub-components of $B$ becomes more like HS, and the concentration of the critical point for $B$ increases with $\Delta_{B_aB_b}$.
		
	In figure \ref{Fig:broader5050DeltaBB} we increased the $PD$ for component $B$ to 8.00 and 12.00 respectively, we kept $\Delta_{AB} = 0$ and we varied $\Delta_{B_aB_b}$ as before. With increased $PD$, the two-phase boundary of the polydisperse mixture shifts towards lower packing fractions for all mixtures. The effect of $\Delta_{B_aB_b}$ on the position of the two-phase boundary becomes smaller at lower concentration of $B$, however at higher concentrations of $B$ we see that the two-phase boundary for positive $\Delta_{B_aB_b}$ bends towards the y-axis and this effect is more pronounced for higher $PD$.
	The polydispersity of $B$ also has an effect on the position of the three-phase boundary. With increased $PD$, the position of the three-phase boundary becomes less dependent on $\Delta_{B_aB_b}$ and the difference in the position of the two-phase boundary and the three-phase boundary increases for the mixtures with $\Delta_{B_aB_b} = 0.10$. For the mixtures with $PD = 12.00$, the difference between the three-phase boundary for the mixtures that phase separate into three phases becomes negligible.
	We see similar trends in the critical points for the more polydisperse mixture as in figure \ref{Fig:narrow5050DeltaBB}, however, with increased polydispersity and especially increased incompatibility between the sub-components ($\Delta_{B_aB_b}>0$), the critical point shifts towards higher concentrations of $B$. For the mixtures with larger $\Delta_{B_aB_b}$, the critical point can shift to $\eta_{B_{crit}} > 0.5$.
	
	In figure \ref{Fig:PD2575} and \ref{Fig:PD1090} we varied the ratio between the sub-components of $B$. The ratio between the sub-components of $B$ is 25/75 with a $PD = 6.93$ in figure \ref{Fig:PD2575} (both left and right skewed) and 10/90 with a $PD = 4.80$ in figure \ref{Fig:PD1090} (both left and right skewed). These mixtures can be seen as a model for mixtures that contain some impurities, from a similiar material but at at different size when $\Delta_{B_aB_b} = 0$ or a material that is less compatible with the main component (when $\Delta_{B_aB_b}>0$) or more compatible with the main component (when $\Delta_{B_aB_b}<0$). The $PD$ is the same for both the left skewed and the right skewed mixtures. For both types of mixtures, we see that the two-phase boundaries are closer to each other for the left-skewed mixtures (large amount of the largest sub-component) compared to the right-skewed mixtures. Next to that, these left-skewed mixtures also show a larger bend in the two-phase boundary towards the y-axis for $\Delta_{B_aB_b} > 0$. The mixture in figure \ref{Fig:Left10} with $\Delta_{B_aB_b} = 0.05$ does not have a three phase boundary, even though mixtures with these sizes can phase separate into three phases: the distribution of the sub-components makes these concentrations unattainable in the range of concentrations we focus on.		
	
	For the right-skewed mixtures, we see that the three-phase boundary for mixtures with $\Delta_{B_aB_b} = 0.10$ comes very close to the two-phase boundary and for mixtures with $\Delta_{B_aB_b} = 0.05$ the three phase boundary shows a bend back towards lower concentrations of $A$ at low concentrations of $B$. This is due to the shape of the three-phase surface and can also be seen on a small level in the mixture $\Delta_{B_aB_b} = 0.10$ in figure \ref{Fig:narrow5050DeltaBB}.
	
	Also Bellier-Castella and coauthors \cite{Bellier-Castella2000} found the possibility of three phase separation for polydisperse components. According to them, the transition between the two phase and three phase region proceeds via a second critical point. This second critical point is polydispersity induced.

	\subsection{\label{ss:DeltaBBDeltaAB} Mixtures with non-additivity between sub-components of  $B$ ($\Delta_{B_aB_b}$), and between $A$ and $B$ ($\Delta_{AB}$)}
	
	\begin{figure}
		\includegraphics[trim=1cm 0cm 1cm .5cm, clip=true ,width=0.47\textwidth]{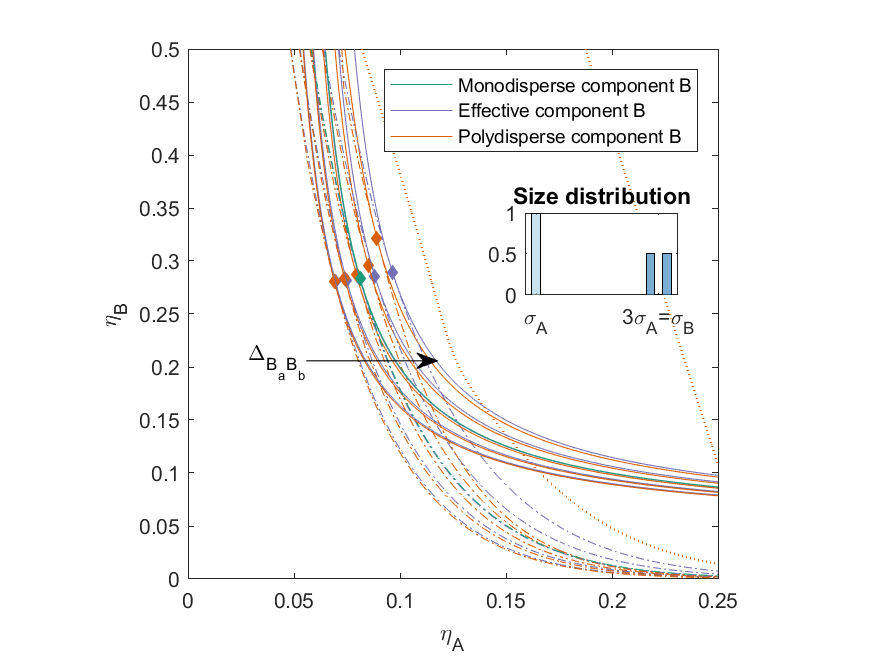}
		\caption{\label{Fig:combined}Phase diagram for binary (component $A$ and $B$) non-additive hard sphere mixture with size ratio $q = \sigma_A / \sigma_B =  1/3$, component $A$ is monodisperse, component $B$ is polydisperse ($PD = 4.00$), plotted as a function of the partial packing fractions, $\eta_A$ and $\eta_B$. The interaction between components $A$ and $B$ is non-additive with a non-additivity parameter $\Delta_{AB} = 0.075$, the interaction between the sub-components $B$ is non-additive, the non-additivity parameter $\Delta_{B_aB_b}$ was varied from -0.1 to 0.1 with a step size of 0.05 (the arrow indicates increasing $\Delta_{B_aB_b}$). The spinodal (solid line) and binodal (dashed line) meet each other at the plait point (diamond), the three phase boundary is indicated with a dotted line.}
	\end{figure}

	In figure \ref{Fig:combined} we plot the phase diagram for mixtures with varying $\Delta_{B_aB_b}$, with a size ratio between component $A$ and $B$ of $q = 1/3$, and a non-additive interaction between $A$ and $B$ $\Delta_{AB} = 0.075$. This is in fact a combination of the case in section \ref{ss:DeltaAB} and \ref{ss:DeltaBB} at lower size ratio between $A$ and $B$. The polydisperisty of $B$ is 4.00 (for mixtures with more variety in $PD$ and $\Delta_{AB}$ we refer to supplementary material). The phase diagram of these mixtures shows a lot of similarities with the phase diagram of the mixtures in figure \ref{Fig:narrow5050DeltaBB} though at different $\eta_A$ due to the different size ratio. Since the mixtures in figure \ref{Fig:narrow5050DeltaBB} have the same $PD$, we conclude that the three-phase boundary position and shape is largely dependent on the interaction between the sub-components of $B$. The interaction between the sub-components is determined by both the $PD$ and the non-additivity parameter $\Delta_{B_aB_b}$.
	
	\makeatletter\onecolumngrid@push\makeatother
	
	\begin{table*}[h]

		\caption{\label{tab:critical}Critical points for the different binary mixtures depending on the non-additivity of component $B$, see also figure \ref{Fig:narrow5050DeltaBB} and phase separated concentrations and volume fraction $\alpha$ of the different mixtures for specific parent concentration ($\eta_{A_{parent}} =0.010, \eta_{B_{parent}} =0.200$), depending on the non-additivity of component $B$ see \ref{tab:fractionhist} for distribution of component $B$ in each phase.}
		\centering

		\begin{tabular}{l l p{4.5cm} p{4.5cm} p{4.5cm}}
			\hline
			\textbf{$ \Delta_{B_aB_b}$} & \textbf{$\eta_{crit}$} &\textbf{Top phase} &\textbf{Middle phase}& \textbf{Bottom phase}\\  
			\hline
			0.100 & (0.007, 0.323) & $\eta$ (0.011, 0.074) & $\eta$ (0.006, 0.596)&   $\eta$ (0.004,0.953)\\
			&  & PD: 3.35, Size: 0.97, $\alpha:$ 0.817 & PD: 3.84,  Size: 0.99, $\alpha:$ 0.097 & PD: 2.62, Size: 1.03, $\alpha:$ 0.087\\
			\hline
			0.0875 & (0.007, 0.303) & $\eta$ (0.011, 0.064) & $\eta$ (0.005, 0.733)&   $\eta$ (0.004, 0.813)\\
			& & PD: 3.39, Size: 0.98, $\alpha:$ 0.804 & PD: 3.94,  Size: 1.01, $\alpha:$ 0.138 & PD: 3.65, Size: 1.02, $\alpha:$ 0.058\\
			\hline
			0.075 & (0.007, 0.294) & $\eta$ (0.011, 0.057) & &   $\eta$ (0.005, 0.767)\\
			& & PD: 3.42, Size: 0.98, $\alpha:$ 0.799 & & PD: 3.91, Size: 1.01, $\alpha:$ 0.201\\
			\hline
			0.05 & (0.007, 0.285) & $\eta$ (0.011, 0.047) & &   $\eta$ (0.004, 0.800)\\
			& & PD: 3.44, Size: 0.98, $\alpha:$ 0.797 & & PD: 3.94, Size: 1.01, $\alpha:$ 0.203\\
			\hline
			0 & (0.007, 0.280) & $\eta$ (0.011, 0.034) & &   $\eta$ (0.004, 0.881)\\
			& & PD: 3.45, Size: 0.98, $\alpha:$ 0.804 & & PD: 3.97, Size: 1.00, $\alpha:$ 0.196\\
			\hline
			-0.05 & (0.006, 0.281) & $\eta$ (0.011, 0.026) & &   $\eta$ (0.004, 0.966)\\
			& & PD: 3.45, Size: 0.98, $\alpha:$ 0.815 & & PD: 3.98, Size: 1.00, $\alpha:$ 0.185\\
			\hline
			.-0.1 & (0.006, 0.285) & $\eta$ (0.011, 0.020) & &   $\eta$ (0.003, 1.051)\\
			& & PD: 3.45, Size: 0.98, $\alpha:$ 0.825 & & PD: 3.99, Size: 1.00, $\alpha:$ 0.175\\
			\hline
		\end{tabular}

	\end{table*}

	\clearpage

	\makeatletter\onecolumngrid@pop\makeatother

	\subsection{\label{ss:Fractionation} Fractionation}
	
	When a parent mixture demixes into two or more phases, each component (and also their sub-components) in the mixture will find its preferential phase in order to minimize the Helmholtz free energy of the system. This leads each phase to be enriched in one of the components, whilst being depleted by the other component(s). The other component(s) are then present only at low volume fractions. We investigated the phase separation for the mixtures in section \ref{ss:DeltaBB} for a specific parent mixture ($\eta_{A_{parent}} = 0.010, \eta_{B_{parent}} = 0.200$) in terms of the volume fraction of both components in each phase, the degree of polydispersity of component $B$, average size of component $B$ in the child phases compared to the average size of component $B$ in the parent phase and the volume fraction of the phases ($\alpha$) see table \ref{tab:critical} for the mixtures from figure \ref{Fig:narrow5050DeltaBB} (mixtures with $PD = 4.00$), for other mixtures we refer to the supplementary materials. The composition histograms for each phase are given in table \ref{tab:fractionhist} for the same mixture, for other mixtures we refer to the supplementary materials. Since at this parent concentration the mixture with non-additivity parameter $\Delta_{B_aB_b} = 0.10$ separates into three phases, we have also calculated the child phases for mixtures with between $\Delta_{B_aB_b} = 0.10$ and $\Delta_{B_aB_b} = 0.05$ to investigate the behavior of the sub-components $B$ depending on the non-additivity. 
	
	For all mixtures, the top phase, which is also the largest phase in volume, is enriched in component $A$. The volume fraction of the top phase is dependent on the non-additive interaction between the sub-components of $B$. It increases with both more compatibility between the sub-components as well as less compatibility, with a minimum volume fraction at $\Delta_{B_aB_b} = 0.05$. We find this dependence in volume fraction on the non-additivity parameter $\Delta_{B_aB_b}$ also for the other mixtures, however the minimum volume fraction is at different $\Delta_{B_aB_b}$ depending on the sizes of and the ratio of the sub-components $a$ and $b$ of $B$. For the mixtures ($\Delta_{B_aB_b} > 0.075$) that phase separate into three phases at this parent mixture concentration, we conclude that it is mostly the bottom phase that demixes into an additional phase (the middle phase). The bottom phase is enriched in the largest sub-component of $B$, while the top phase (and middle phase to a lesser extent) is enriched in the smaller sub-component of $B$. We see this behavior also for the other mixtures with different composition of $B$.

	\makeatletter\onecolumngrid@push\makeatother

	\begin{table*}

		\caption{\label{tab:fractionhist}Phase separation of different mixtures and fractionation of component $B$ for specific parent distribution ($\eta_{A_{parent}} =0.010, \eta_{B_{parent}} =0.200$), depending on the non-additivity of component $B$, see also figure \ref{Fig:narrow5050DeltaBB}}
		\centering
		\begin{tabular}{p{5 cm} c c c c}
			\hline
			\multicolumn{1}{l}{\textbf{$ \Delta_{B_aB_b}$}}&
			\multicolumn{1}{c}{\textbf{Parent distribution}}&
			\multicolumn{1}{c}{\textbf{Top phase}}&
			\multicolumn{1}{c}{\textbf{Middle phase}}&
			\multicolumn{1}{c}{\textbf{Bottom phase}}\\  
			\hline
			0.1 &\adjustbox{valign=t}{\includegraphics[trim = 1.3cm 6cm 9.7cm 3cm,clip=true, width= 2.5 cm]{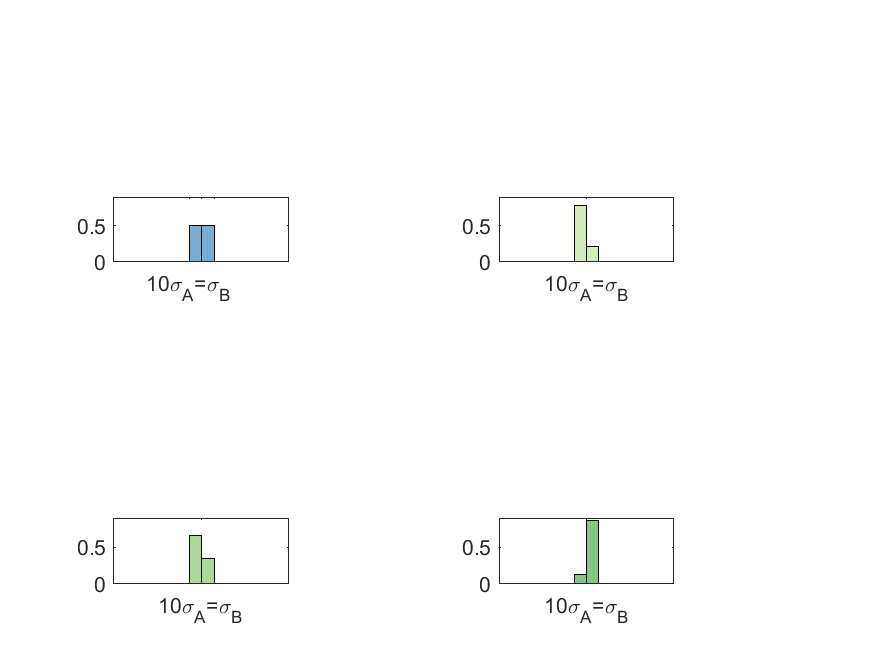}}  & \adjustbox{valign=t}{\includegraphics[trim = 7.8cm 6cm 3.2cm 3cm,clip=true, width= 2.5 cm]{PDmix_histograms_AB_AB_25_i2_q10_delta010.png}} & \adjustbox{valign=t}{\includegraphics[trim = 1.3cm .5cm 9.7cm 8.5cm,clip=true, width= 2.5 cm]{PDmix_histograms_AB_AB_25_i2_q10_delta010.png}} & \adjustbox{valign=t}{\includegraphics[trim = 7.8cm .5cm 3.2cm 8.5cm,clip=true, width= 2.5 cm]{PDmix_histograms_AB_AB_25_i2_q10_delta010.png}} \\
			0.0875 &\adjustbox{valign=t}{\includegraphics[trim = 1.3cm 6cm 9.7cm 3cm,clip=true, width= 2.5 cm]{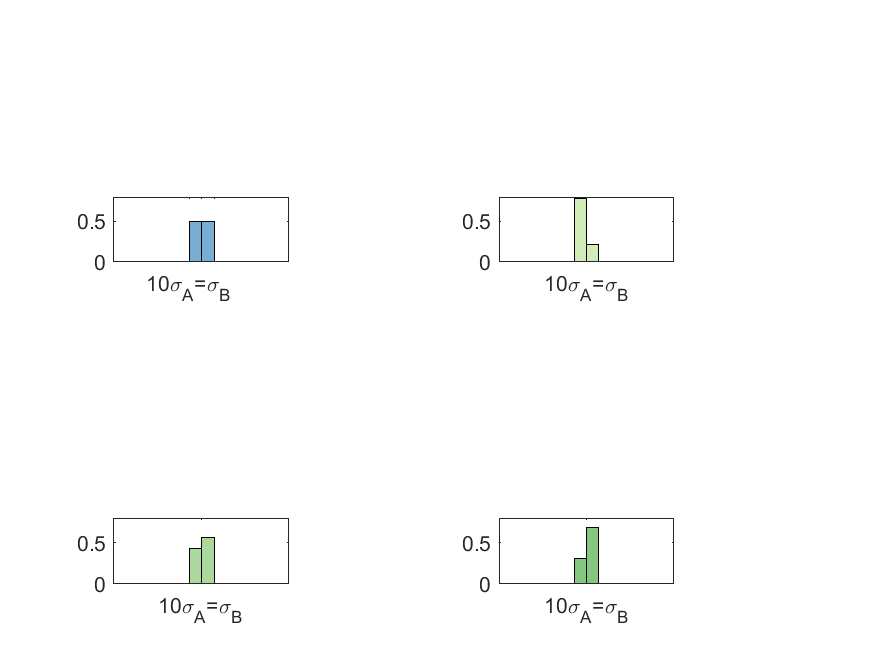}}  & \adjustbox{valign=t}{\includegraphics[trim = 7.8cm 6cm 3.2cm 3cm,clip=true, width= 2.5 cm]{PDmix_histograms_AB_AB_25_i2_q10_delta09.png}} & \adjustbox{valign=t}{\includegraphics[trim = 1.3cm .5cm 9.7cm 8.5cm,clip=true, width= 2.5 cm]{PDmix_histograms_AB_AB_25_i2_q10_delta09.png}} & \adjustbox{valign=t}{\includegraphics[trim = 7.8cm .5cm 3.2cm 8.5cm,clip=true, width= 2.5 cm]{PDmix_histograms_AB_AB_25_i2_q10_delta09.png}} \\
			0.075 &\adjustbox{valign=t}{\includegraphics[trim = 1.3cm .5cm 9.7cm 8.5cm,clip=true, width= 2.5 cm]{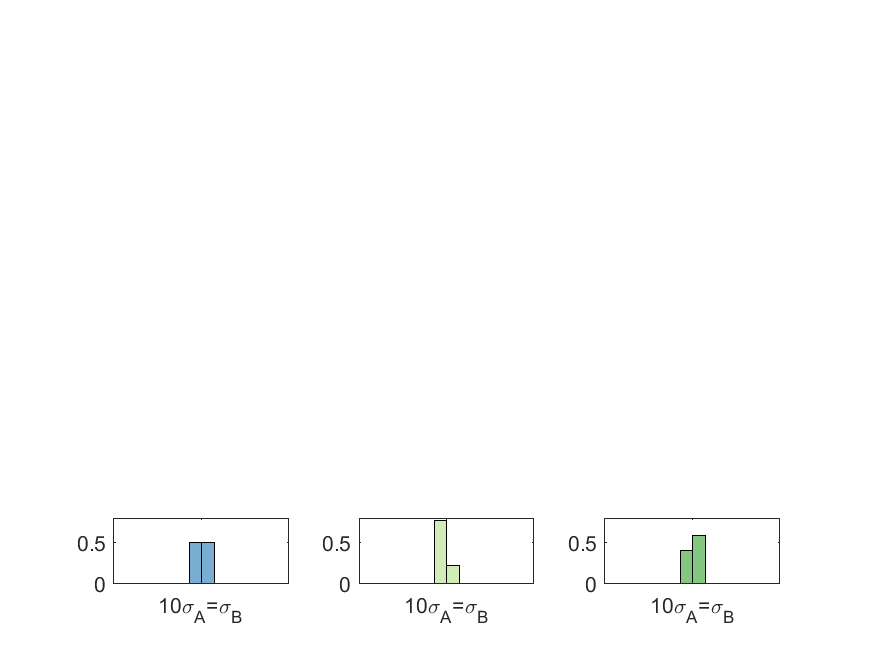}}  & \adjustbox{valign=t}{\includegraphics[trim = 5.3cm .5cm 5.7cm 8.5cm,clip=true, width= 2.5 cm]{PDmix_histograms_AB_AB_25_i2_q10_delta08.png}}& &\adjustbox{valign=t}{\includegraphics[trim = 9.5cm .5cm 1.5cm 8.5cm,clip=true, width= 2.5 cm]{PDmix_histograms_AB_AB_25_i2_q10_delta08.png}} \\
			0.05 &\adjustbox{valign=t}{\includegraphics[trim = 1.3cm .5cm 9.7cm 8.5cm,clip=true, width= 2.5 cm]{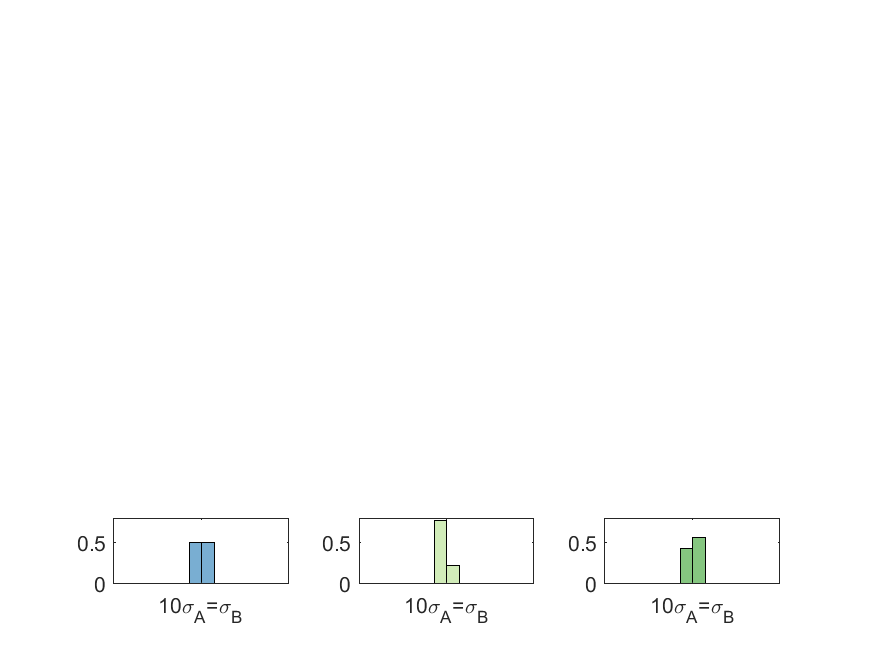}}  & \adjustbox{valign=t}{\includegraphics[trim = 5.3cm .5cm 5.7cm 8.5cm,clip=true, width= 2.5 cm]{PDmix_histograms_AB_AB_25_i2_q10_delta05.png}}& &\adjustbox{valign=t}{\includegraphics[trim = 9.5cm .5cm 1.5cm 8.5cm,clip=true, width= 2.5 cm]{PDmix_histograms_AB_AB_25_i2_q10_delta05.png}} \\
			0 &\adjustbox{valign=t}{\includegraphics[trim = 1.3cm .5cm 9.7cm 8.5cm,clip=true, width= 2.5 cm]{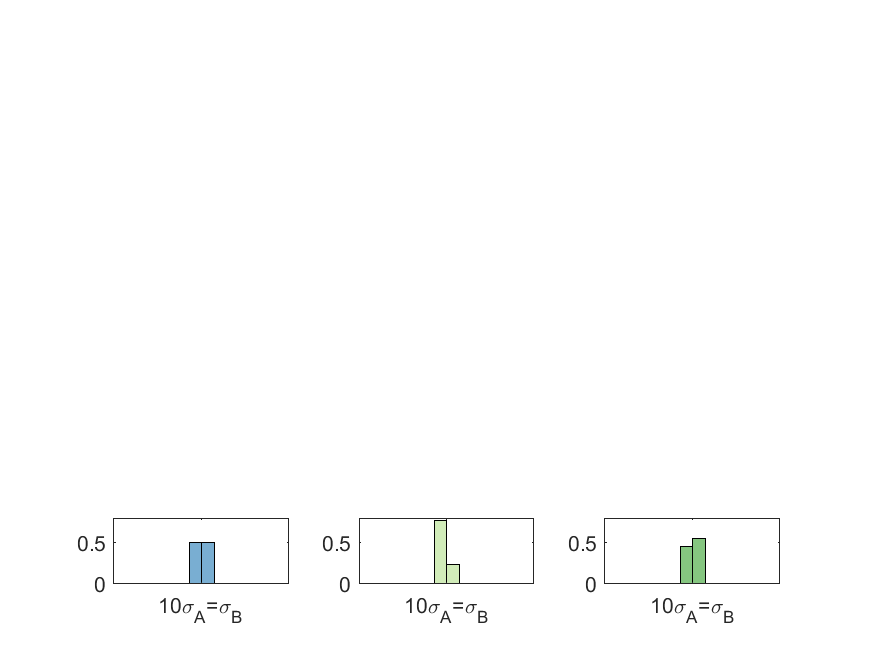}}  & \adjustbox{valign=t}{\includegraphics[trim = 5.3cm .5cm 5.7cm 8.5cm,clip=true, width= 2.5 cm]{PDmix_histograms_AB_AB_25_i2_q10_delta00.png}}& &\adjustbox{valign=t}{\includegraphics[trim = 9.5cm .5cm 1.5cm 8.5cm,clip=true, width= 2.5 cm]{PDmix_histograms_AB_AB_25_i2_q10_delta00.png}} \\
			-0.05 &\adjustbox{valign=t}{\includegraphics[trim = 1.3cm .5cm 9.7cm 8.5cm,clip=true, width= 2.5 cm]{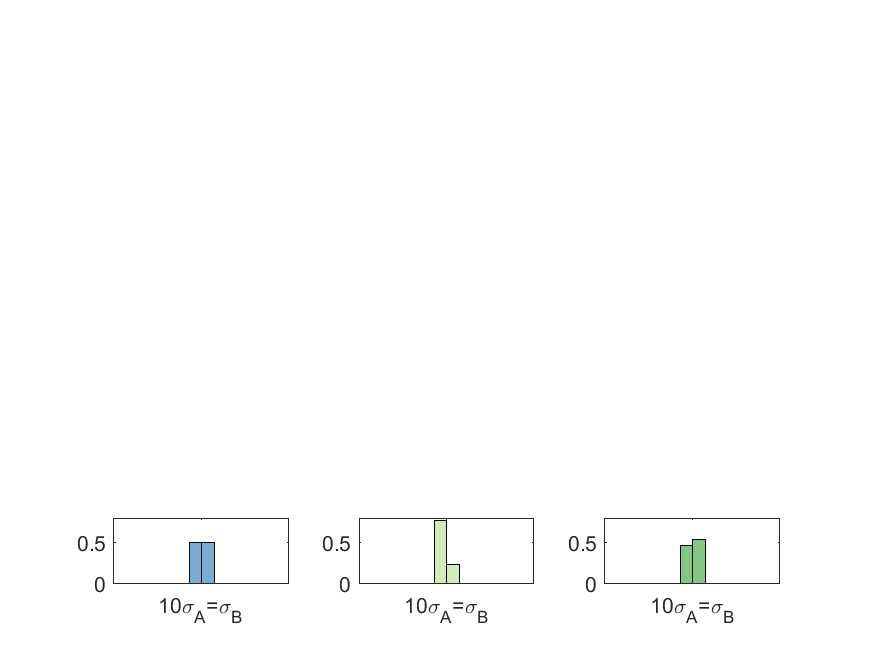}}  & \adjustbox{valign=t}{\includegraphics[trim = 5.3cm .5cm 5.7cm 8.5cm,clip=true, width= 2.5 cm]{PDmix_histograms_AB_AB_25_i2_q10_delta0-5.png}}& &\adjustbox{valign=t}{\includegraphics[trim = 9.5cm .5cm 1.5cm 8.5cm,clip=true, width= 2.5 cm]{PDmix_histograms_AB_AB_25_i2_q10_delta0-5.png}} \\
			-0.1 &\adjustbox{valign=t}{\includegraphics[trim = 1.3cm .5cm 9.7cm 8.5cm,clip=true, width= 2.5 cm]{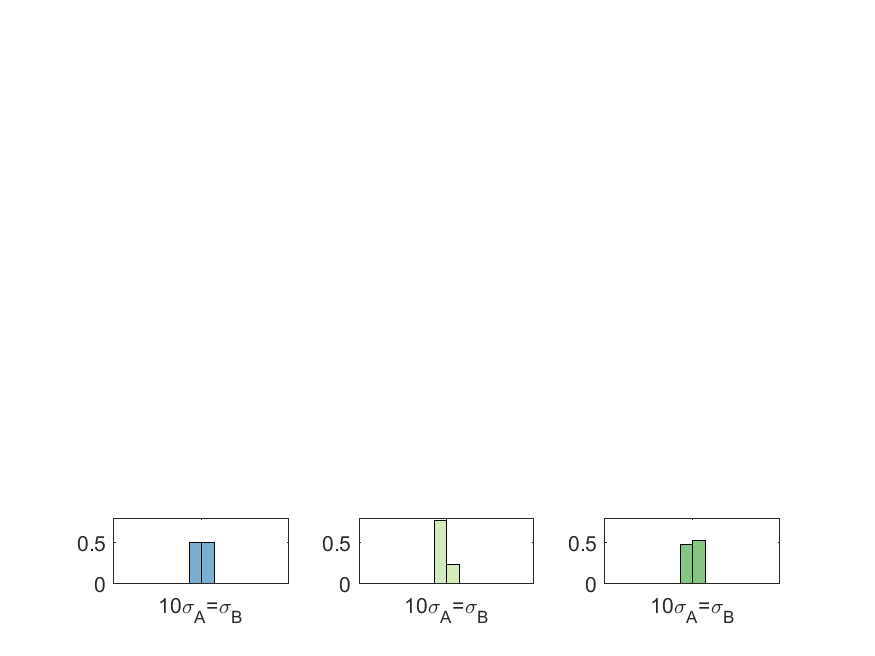}}  & \adjustbox{valign=t}{\includegraphics[trim = 5.3cm .5cm 5.7cm 8.5cm,clip=true, width= 2.5 cm]{PDmix_histograms_AB_AB_25_i2_q10_delta0-10.png}}& &\adjustbox{valign=t}{\includegraphics[trim = 9.5cm .5cm 1.5cm 8.5cm,clip=true, width= 2.5 cm]{PDmix_histograms_AB_AB_25_i2_q10_delta0-10.png}} \\
			\hline
		\end{tabular}

	\end{table*}

	\clearpage	
	\makeatletter\onecolumngrid@pop\makeatother

	The fractionation of the sub-components of $B$ is dependent on the non-additivity parameter $\Delta_{B_aB_b}$. When $\Delta_{B_aB_b} <0 $ the sub-components $a$ and $b$ are more compatible with each other and this leads to less fractionation, as can be seen in table \ref{tab:fractionhist}, while on the other hand  $\Delta_{B_aB_b} >0 $ the sub-components are less compatible with each other and more fractionation occurs, even leading to additional phase separation at higher $\Delta_{B_aB_b}$. This is something we see also for the other mixtures (see supplementary material).

	\section{Conclusion}
	
	We find that when the compatibility between component $A$ and $B$ is decreased, the phase diagram (the critical point, phase boundary and spinodal) shifts towards lower volume fractions. This is in line with literature on the phase behavior of NAHS binary monodisperse mixtures. The interaction between $A$ and $B$ is driven by the size ratio ($q$) between $A$ and $B$ and the non-additivity parameter $\Delta_{AB}$.
	
	When the compatibility between the sub-components of the polydisperse component $B$ is altered, the phase diagram changes more drastically. When the compatibility between the sub-components is decreased, the mixture can demix into three phases each enriched in one of the (sub)components of the parent mixture. The shape and position of the three phase boundary is mainly dependent on the interactions between the sub-components of $B$. This means that it is dependent on the non-additivity parameter ($\Delta_{B_aB_b}$) as well as the size ratios and distribution of the sub-components (the degree of polydispersity $PD$). Next to that, depending on the size ratios and distribution of the sub-components we see also that the the binodal and spinodal bend towards the y-axis for higher volume fractions of $B$ when $\Delta_{B_aB_b}$ increases. For the mixtures with a more pronounced bend in the phase boundary and spinodal, we find that the critical point shifts to volume fractions $\eta_{B_{crit}}>0.5$. This behavior is driven to a large extent by the the non-additivity parameter ($\Delta_{B_aB_b}$) as well as the size ratios and distribution of the sub-components (the degree of polydispersity $PD$) and to a lesser extent by the interaction between $A$ and $B$. When the compatibility between the sub-components is increased, the mixture demixes at slightly lower packing fractions compared to the HS mixture. The fractionation of the polydisperse sub-components of $B$ is also dependent on the non-additivity parameter $\Delta_{B_aB_b}$. Less fractionation occurs when  $\Delta_{B_aB_b} <0 $, more fractionation occurs when $\Delta_{B_aB_b} >0 $. At higher $\Delta_{B_aB_b}$ this can even lead to additional phase separation, creating a third phase.

	The virial coefficient approach for polydisperse mixtures allows for the prediction of the phase behavior of polydisperse or impure binary mixtures. Not only does it allow for plotting the phase diagram, it also allows for the calculation of the composition and fractionation of each component in each phase.
	
	\printnomenclature
	\bibliographystyle{unsrt} 
	\bibliography{PolydispersityLit}
\end{document}